# Adsorption-Driven Symmetry Lowering in Single Molecules Revealed by Ångstrom-scale Tip-Enhanced Raman Imaging

Rodrigo Cezar de Campos Ferreira[1,2 ‡], Borja Cirera[3 ‡], Jiří Doležal[1], Álvaro Gallego de Roa[4], Amandeep Sagwal[2,5], Petr Kahan[2], Rubén Canales[3], Fernando Aguilar-Galindo[4,6]*, Martin Švec[1,2]*, and Pablo Merino[3]*

[1] Institute of Organic Chemistry and Biochemistry, Czech Academy of Sciences, Praha 6 CZ16000, Czech Republic

[2] Institute of Physics, Czech Academy of Sciences, Praha 6 CZ16200, Czech Republic

[3] Instituto de Ciencia de Materiales de Madrid, Madrid ES28049, Spain

[4] Departamento de Química, Universidad Autónoma de Madrid, Madrid ES28049, Spain

[5] Faculty of Mathematics and Physics, Charles University; Ke Karlovu 3, CZ12116 Praha 2, Czech Republic

[6] Institute for Advanced Research in Chemical Sciences (IAdChem), Universidad Autónoma de Madrid, Madrid ES28049, Spain

‡ authors contributed equally

Email: fernando.aguilar-galindo@uam.es

Email: svec@fzu.cz

Email: pablo.merino@csic.es

**Abstract**:

The vibrational landscape of adsorbed molecules is central to understanding surface interactions at the atomic scale, influencing phenomena from catalysis to molecular electronics. Recent advances in atomic-scale tip-enhanced Raman spectroscopy (TERS) have enabled vibrational mapping of single molecules with sub-nanometer spatial resolution, providing unprecedented insights into molecule-surface interactions by confining light in plasmonic picocavities. Here, we exploit TERS in a cryogenic scanning tunneling microscope junction to perform Raman hyperspectral mapping of single iron phthalocyanine (FePc) molecules in three non-equivalent adsorption configurations on Ag surfaces. We explore the changes in the vibrational modes of FePc molecules adsorbed on two distinct silver crystal terminations with differing symmetry, Ag(111) and Ag(110), revealing how subtle variations in the adsorption geometry due to substrate anisotropy can strongly influence molecular vibrations, lifting the degeneracy of individual normal modes. Our findings not only demonstrate the first use of sub-nanometer TERS mapping across different symmetry configurations but also provide a deeper understanding of how site-specific vibrational properties are intimately linked to local atomic environments. This capability paves the way for precisely tailoring surface interactions and controlling chemical reactions at the atomic scale.

Molecular symmetry imposes strict selection rules on optical transitions because the interaction between light and matter must obey the symmetry of the molecule's electronic, vibrational, and rotational states.[1] In particular, Raman activity depends on how the polarizability tensor changes during a vibration; a vibrational mode is Raman-active if its irreducible representation matches a linear combination of the polarizability tensor components for the molecule's point group. However, exposing molecules to specific local nanoenvironments can result in atomic structure relaxation leading to reduced symmetry.[2,3] This can have a profound effect on the scattering activities of particular vibrational modes, which may become partially allowed, forbidden or become non-degenerate.[4] Currently, very little is known about the effect of atomic deformations on Raman activity at the scale of single molecules.[5,6]

Tip-enhanced Raman spectroscopy (TERS) in cryogenic conditions within atomically defined plasmonic picocavities[7] has recently demonstrated the ability to visualize individual vibrational modes in real space with sub-molecular resolution.[8-11] With this technique it is possible to reconstruct the chemical structure of single molecular adsorbates, reaching single bond limit,[12-14] track formation and breaking of individual chemical bond in the real time,[15,16] or even sequence fragments of single DNA strands.[17,18] TERS also permits to perform correlative measurements that cannot be tackled with conventional surface-enhanced Raman spectroscopy, e.g. it allows exploring the links between the electronic configurations of individual adsorbates and their Raman spectra upon precise molecular nanomanipulations [19] or controlling vibrations with time-resolved schemes at fs temporal resolution.[20] In addition, evidence from STM experiments shows that single molecules on metal surfaces can undergo structural modifications [21,22] and Jahn-Teller distortions upon molecular adsorption and concomitant spontaneous charging.[23,24] However, the role of distortions upon atomic-scale interactions across different interfaces has not been addressed in TERS experiments with submolecular resolution.

Here we explore the breakdown of the $D_{4h}$ symmetry of iron phthalocyanine (FePc) upon adsorption on the Ag(111) and Ag(110) surfaces using hyperspectral TERS mapping (schematically shown in Fig. 1a and described in methods and Fig. S1). We study three distinct adsorption configurations with progressively lowered $C_2$, $C_{2v}$ and $C_{2d}$ symmetries ($C_{2d}$ being a non-equivalent $C_{2v}$ symmetry with the mirror planes lying on the diagonals of the molecule). We directly observe the symmetry reduction of the vibrations in the real-space-resolved TERS intensity patterns, evidencing the degeneracy lifting in individual normal modes. We confront our experimental findings with density functional theory (DFT) calculations and corroborate the link between the molecular adsorption geometry, reduction of the symmetry and the consequent alterations of the spatio-spectral Raman fingerprints.

Figure 1a shows a STM image of submonolayer FePc adsorbed on Ag(111) with a characteristic four-lobe and bright center contrast.[25] FePc molecules are dispersed on the surface with three equivalent orientations, with one of the main molecular axes (defined by N-Fe-N directions) aligned with any of the three crystallographic low index $\langle 1\bar{1}0 \rangle$ directions of the substrate. Alternatively, FePc adsorbed on Ag(110) adopts two non-equivalent configurations: either with the N-Fe-N axes at 45° along the main crystallographic directions (molecules marked with blue crosses in Fig. 1c), hereafter referred as *aligned FePc/Ag(110)*, or rotated approximately ±30° with respect to the $[1\bar{1}0]$ crystal directions (purple crosses), referred to as *rotated FePc/Ag(110)* hereafter. Due to the symmetry of the molecule-substrate system, the rotated FePc/Ag(110) is found with its chiral equivalent on the surface (see Fig. S2 in Supporting Information).

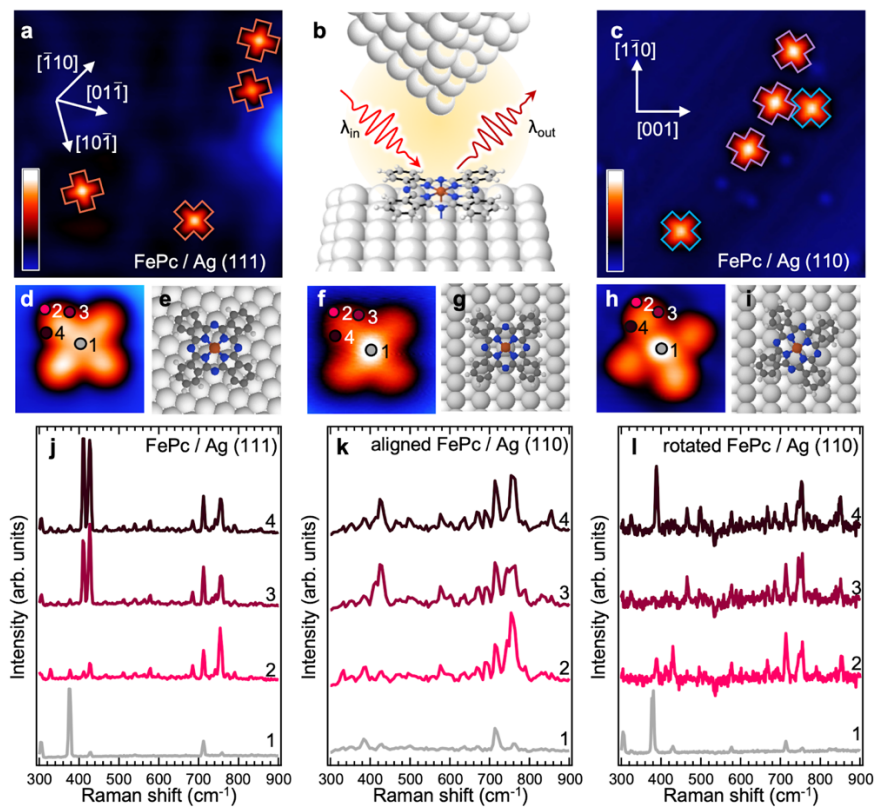

*Figure 1. STM topography imaging of the FePc adsorption configurations on Ag surfaces and TERS point spectra. a. STM image of FePc adsorbed on Ag(111). Main crystallographic directions of the surface are marked with white arrows. The molecules in three equivalent adsorption geometries are marked with orange crosses. Size 15 x 15 nm$^2$, setpoint 50 mV, 30 pA. Scalebar indicates topographic height from 0 to 1.4 Å. b. Experimental scheme: a laser-illuminated plasmonic tip scans above a FePc molecule on Ag(110) inducing Raman scattering of the light on the molecular vibrational modes. c. STM image of FePc adsorbed on Ag(110). The aligned and rotated adsorption configurations are denoted with blue and purple crosses, respectively. Size 15 x 15 nm$^2$, 100 mV, 16 pA. Scalebar indicates topographic height from 0 to 1.4 Å. d. Detailed STM image of FePc/Ag(111). Colored circles point to the position where TERS spectra have been measured. Size 2.0 x 2.0 nm$^2$, setpoint 1 mV, 200 pA. e. Ball-and-stick scheme of FePc/Ag(111) adsorption configuration. f High resolution STM image of aligned FePc/Ag(110). Size 2.0 x 2.0 nm$^2$, setpoint 1 mV, 200 pA. h High resolution STM image of rotated FePc/Ag(110). Size 2.0 x 2.0 nm$^2$, setpoint 1 mV, 1 nA. i. Ball-and-stick scheme of rotated FePc/Ag(110) adsorption configuration. j, k, l. Comparison of the spectra taken on FePc on different intramolecular locations color coded in panels d, f and h, respectively. The spectra are shifted vertically for clarity. Acquisition parameters: 50 s, 1 mV, 200 pA.*

For each adsorption configuration we measure point TERS spectra of the low wavenumber fingerprint region at four positions; three positions at one lobe and one in the center (marked by colored points and numbers in Fig. 1d,f,h). The spectra are presented in Fig. 1 j,k,l for FePc/Ag(111), aligned FePc/Ag(110) and rotated FePc/Ag(110), respectively. We notice that for each adsorption geometry the spectra have comparable vibrational mode frequencies, but the relative Raman intensities of the bands strongly vary depending on the location of the measurement above the molecule. Interestingly, for the rotated FePc/Ag(110) system, the TERS spectra obtained on the two opposite peripheral arms of the molecule (locations 3 and 4 in Fig. 1h) show significant variations in the relative intensities of the vibrational modes. In contrast, the FePc/Ag(111) and aligned FePc/Ag(110) show nearly identical spectra at these locations. Altogether, these results indicate an adsorption-driven non-homogeneous spatial distribution of the TERS intensity for individual vibrations within the molecules.

To resolve the submolecular intensity distribution in each FePc configuration, we map the TERS signal as a function of tip position. Real-space, band-resolved TERS mapping pinpoint individual modes within molecules.[12, 26-28] The tip-sample height is decisive to maximize the signal to noise ratio. Heights below 50 pm may result in the formation of molecular point contact (see Fig. S3 in Supporting information)[29,30] drastically enhancing the signal but hindering map acquisition. Thus, in order to avoid the perturbation of the picocavity, our hyperspectral maps are performed at a typical tip-height of 100 pm. The Raman spectra generated by adding all point spectra for a given map are shown in Fig 2 a,d,g. From the intensity peaks we can identify the representative subset of modes which are TERS-active for all the three adsorption geometries. We create the spatial maps of these modes by integrating the spectral intensities in specific wavenumber windows in Fig. 2 c, f, i (the full sets of observed band-resolved TERS maps can be found in Fig. S4-7 of the Supplementary information). Each map shows a distinctive pattern with a high level of intramolecular detail, reaching a resolution of 1.6 Å FWHM for the 374 cm$^{-1}$ mode on Ag(111) (see Fig. S8 in Supporting Information). Furthermore, we also distinguish significant variations of intensity distributions in the maps for the same vibrational bands (identified based on similarity of wavenumbers) depending on the adsorption configurations. Surprisingly, we find in many cases that the characteristic mode patterns have a symmetry lower than the expected D$_{4h}$ of the unperturbed molecules e.g. the band around 745 cm$^{-1}$ in all three cases. In the following we link the symmetries of the TERS patterns to the anisotropic effect of the surface in each particular adsorption geometry.

Upon a careful inspection of the FePc/Ag(111) TERS maps in Fig.2c, we see indications of breaking the D$_{4h}$ rotational symmetry for the 424, 747 and 853 cm$^{-1}$ modes. The apparent symmetry in this case defined by two mirror planes along the N-Fe-N axes ($\sigma_d$ and $\sigma'_d$ in the simplified model in Fig. 2b) and a twofold rotation (C$_2$ in Fig. 2b), indicating an overall C$_{2d}$ point symmetry. The $\sigma_d$ and $\sigma'_d$ mirror axes are aligned with the $[1\bar{1}0]$ and $[\bar{2}11]$ surface directions, respectively, coincidental with the two types of mirror planes of the top Ag(111) atomic layer and the main anisotropic directions. For the aligned FePc/Ag(110), the picture is different; the majority of the patterns shown in Fig. 2f also manifest a twofold symmetry with two mirror planes ($\sigma_v$ and $\sigma'_v$ in Fig.2e) now coinciding with the $[1\bar{1}0]$ and $[001]$ main crystallographic directions of the Ag(110) indicating a C$_{2v}$ symmetry. Finally, many of the presented TERS maps of the rotated FePc/Ag(110) in Fig. 2i clearly have a reduced C$_2$ twofold rotational symmetry, in agreement with the adsorption model in Fig. 2h in which the molecular axes are not aligned with any of the principal crystallographic directions of the surface.

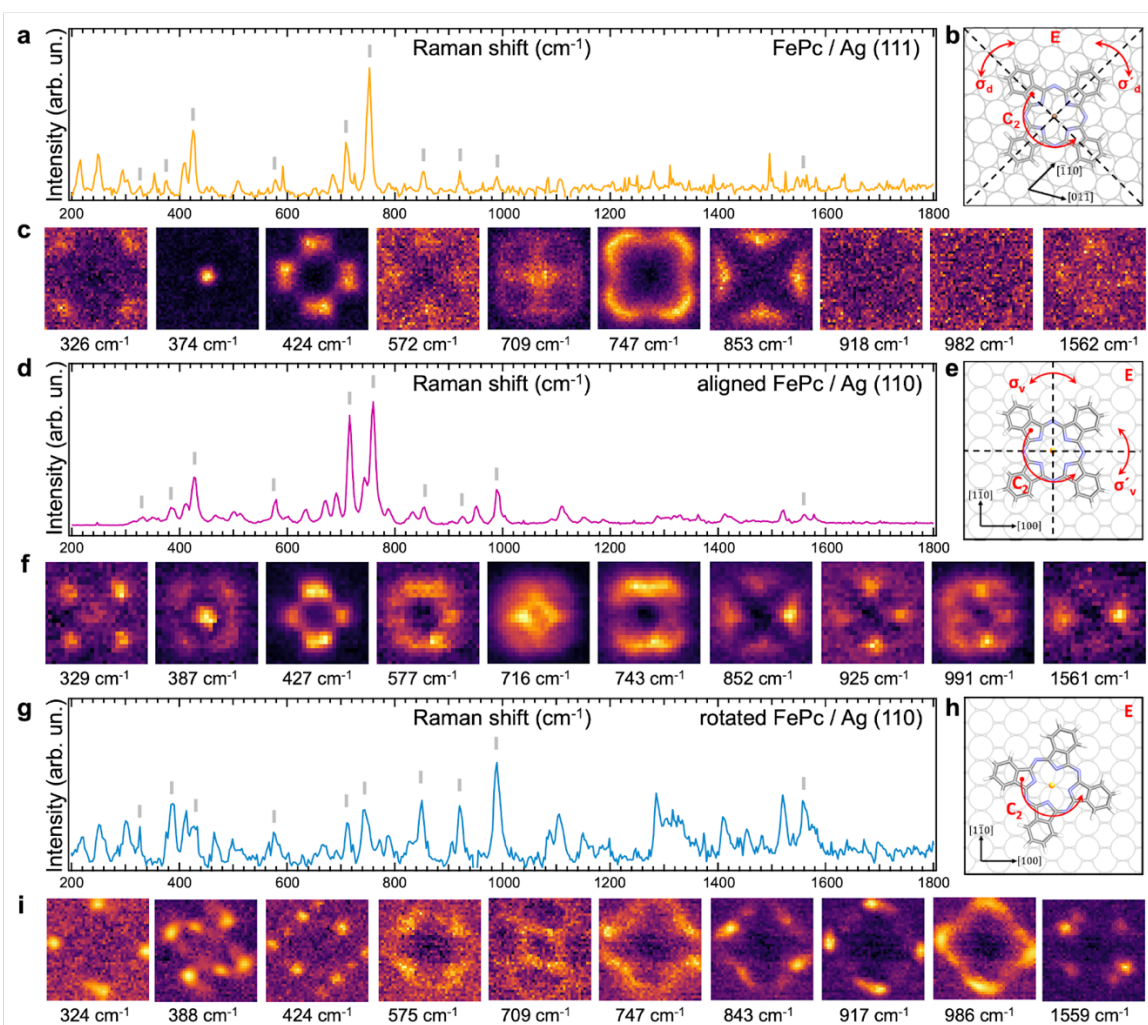

*Figure 2.* TERS hyperspectral maps of FePc on Ag(111) and (110) terminations. *a*. Spatially integrated TERS spectrum of the FePc/Ag(111) *b*. Scheme of the symmetry elements of the FePc/Ag(111) system: E - identity operator. $C_2$ - 180° rotation around the Fe center. $\sigma_d$ and $\sigma'_d$ mirror planes across the diagonals. *c*. Real-space maps of high intensity vibrations of the FePc/Ag(111). *d*. Spatially integrated TERS spectrum of the aligned FePc/Ag(110). *e*. Scheme of the symmetry elements of the aligned FePc/Ag(110) system: E - identity operator. $C_2$ - 180° rotation around Fe center. $\sigma_v$ and $\sigma'_v$ mirror planes across the vertical and horizontal directions, respectively. *f*. Real-space maps of high intensity vibrations of the aligned FePc/Ag(110). *g*. Spatially integrated TERS spectrum of the FePc/Ag(111). *h*. Scheme of the symmetry elements of the rotated FePc/Ag(110) configuration: E - identity operator. $C_2$ - 180° rotation around the Fe center. *i*. Real-space maps of high intensity vibrations of FePc/Ag(111).

FePc has N=57 atoms and therefore 3N - 6 = 165 distinct vibrational modes. In gas phase it is represented by the $D_{4h}$ group irreducible representation with Γ = 14 $A_{1g}$ + 13$A_{2g}$ + 6$A_{1u}$ + 8$A_{2u}$ +14$B_{1g}$

+ 14$B_{2g}$ + 7$B_{1u}$ +7$B_{2u}$ +26$E_g$ + 56$E_u$. The modes $A_{1g}$, $B_{1g}$, $B_{2g}$ and $E_g$ are modes where the molecular vibrations induce a change in the polarizability and therefore are in principle Raman active modes. When lowering the symmetry to $C_{2d}$ and $C_{2v}$, the vibrational modes fall into four irreducible representations $A_1$, $A_2$, $B_1$, and $B_2$; upon further reduction to $C_2$ symmetry, the only two irreducible representations left are A and B modes; all of which can be in principle Raman active.[2] Thus, we expect that the anisotropic relaxation of the molecule in the reduced-symmetry adsorption geometries will result in the appearance of new Raman active modes and the splitting of *D*$_{4h}$ Raman active degenerate modes ($E_g$ modes) in the TERS spectra. We have examined the splitting of the bands in our experiments and found characteristic examples for high intensity split vibrational bands for the three adsorption configurations (shown in Fig.3).

In Fig. 3 a,d,g we present vibrations in the range 380-450 cm$^{-1}$ for the three adsorption configurations which we attribute to two $D_{4h}$ intrinsically degenerate out-of-plane displacements of the atoms belonging to the benzene units (the two degenerate $E_g$ modes at 419 cm$^{-1}$ shown in Fig. S9 of the Supporting Information), according to our gas-phase time-dependent DFT calculations. The second column (Fig. 3 b,e,h) presents the vibrational band 720-780 cm$^{-1}$ corresponding to $E_g$ modes at 764 cm$^{-1}$ in Fig. S9. The third column (Fig. 3 c,f,i) shows the region 800-880 cm$^{-1}$ related to out-of-plane deformation of benzene rings ($E_g$ modes at 872 cm$^{-1}$ in Fig. S9). For FePc in $C_{2d}$ and $C_{2v}$ configurations we are able to resolve two distinct TERS maps with an orthogonal contrast along one of the two mirror planes within a few tens of wavenumbers. This is a clear indication that these doublets come from vibrational modes whose original degeneracy has been lifted.[5] The value of the frequency split of each pair of degenerate vibrations is proportional to the magnitude of the anisotropy of the local potential surface and correlates with the amplitude of the adsorption-induced structural deformation and the anisotropic redistribution of charge and hybridization with the substrate, which modifies the Raman polarizability. We also observe minor shifts of few cm$^{-1}$ for the modes from one adsorption configuration to another, which can be related to vibrational softening/stiffening.[22] For molecules in $C_2$ configuration, the stronger symmetry reduction produces complex chiral patterns in the hyperspectral maps of the split modes and weaker Raman intensities. Overall, we attribute the band splitting and shifts of the originally degenerated Raman active $E_g$ modes to the atomic distortion and concomitant molecular orbital hybridization occurring in response to the interaction with the metal substrate. In Fig. S10 we introduce the theoretical atomic displacements of calculated frequencies presented in Fig. 3. This notion is further supported by the comparison with TERS spectra measured on FePc molecules adsorbed on two monolayers of sodium chloride (NaCl) maintaining a $D_{4h}$ symmetry, where vibrational splitting (degeneracy lifting) is not observed (see Fig. S11 in the supplementary information).

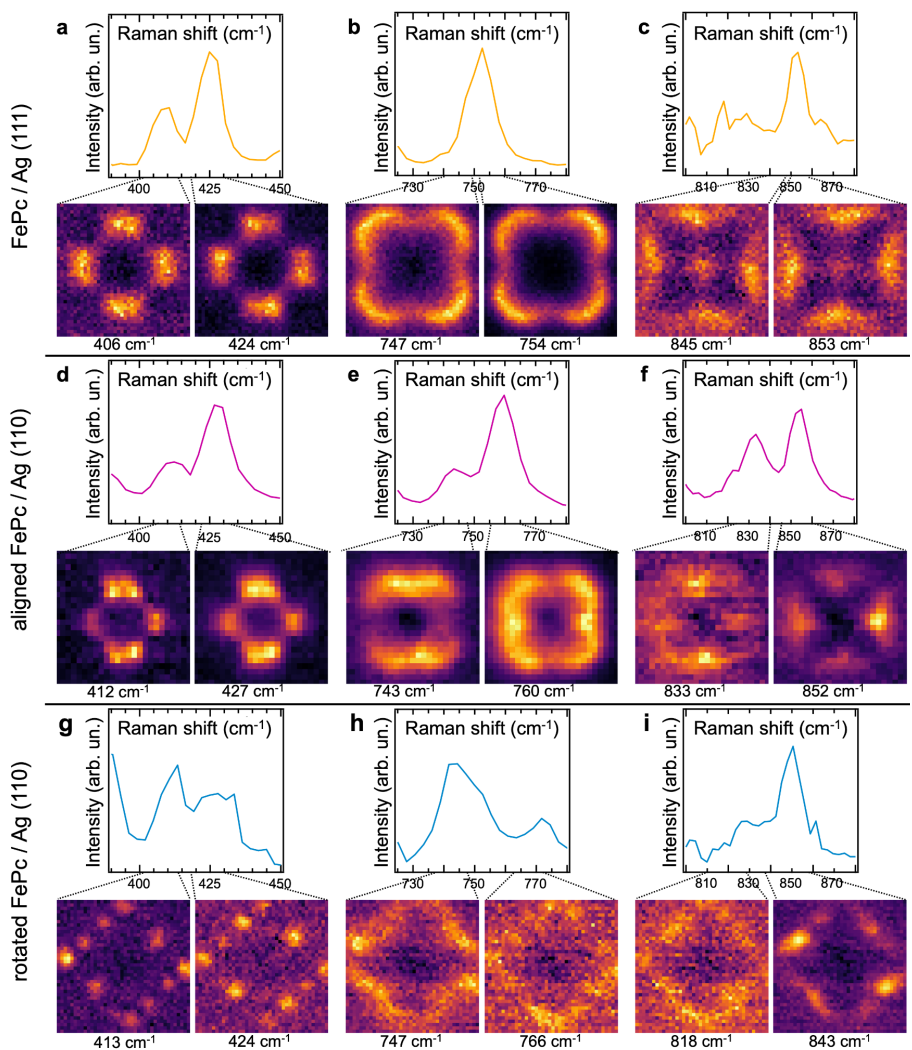

*Figure 3.* Substrate-induced symmetry-lowering and splitting of vibrational modes. ***a***. Experimentally observed TERS doublet in the 380 cm$^{-1}$ to 450 cm$^{-1}$ region of the FePc/Ag(111) system. The maps corresponding to the Raman intensity of each of the two sub-peaks are presented in the lower part. ***b***. TERS peaks in the 720-780 cm$^{-1}$ region of the FePc/Ag(111) and intensity maps of the high and low frequency regions of the peak. ***c***. TERS peaks in the 800-880 cm$^{-1}$ region of the FePc/Ag(111) and intensity maps. ***d***. TERS doublet in the 380-450 cm$^{-1}$ region of the aligned FePc/Ag(111) configuration and intensity maps. ***e***. TERS doublet in the 720-780 cm$^{-1}$ region of the aligned FePc/Ag(111) configuration and intensity maps. ***f***. TERS doublet in the 800-880 cm$^{-1}$ region of the aligned FePc/Ag(111) configuration and intensity maps. ***g***. TERS doublet in the 380-450 cm$^{-1}$ region of the rotated FePc/Ag(111) configuration and intensity maps. ***h***. TERS doublet in the 720-780 cm$^{-1}$ region of the rotated FePc/Ag(111) configuration and intensity maps. ***i***. TERS doublet in the 800-880 cm$^{-1}$ region of the rotated FePc/Ag(111) configuration and intensity maps.

To confirm our interpretation, we have carried out DFT calculations of the different adsorption geometries of FePc on the Ag(111) and Ag(110) surfaces. The lowest energy calculated configurations reproduce the experimentally observed adsorption configurations (see Fig. S12 and S11 in Supporting Information) in agreement with previous experimental results and calculations.[31,32] In Fig.4 we introduce the schematics of the atomic positions for each configuration, with the side views being aligned with high density symmetry directions of the surface. The individual atoms color-coded according to their out-of-plane height coordinate. FePc/Ag(111) adopts a bowl-shaped configuration having the center Fe atom 71 pm lower than the highest H atom of the benzene extrema. Aligned FePc/Ag(110) adopts a saddle shape geometry where two N atoms are the closest to the surface and H atoms on the orthogonal benzene rings lying 59 pm higher. At last, rotated FePc/Ag(110) adopts a configuration where all four benzene rings are distorted to form a propeller-like shape, with the difference between the lowest and highest atoms being 68 pm. Calculations thus confirm that the molecules break their original $D_{4h}$ symmetry and relax to lower symmetry configuration due to the interaction with the surface. The analysis of the symmetry of the optimized structures confirms the new molecular point groups (i.e. $C_{2d}$, $C_{2v}$, $C_2$). We also note that strong charge transfers from the surface to the molecule of ca. 0.8 e$^-$ are predicted, indicating that FePc tends to spontaneously charge on the surface. While the total donated charge is similar, the spatial distribution of that charge is markedly different (see Fig. S13 in Supporting Information). The charge distribution in FePc/Ag(111) is primarily centered around the Fe metal atom, first neighbouring N atoms, and second neighbour carbon atoms, leaving the four benzene extrema electronically unperturbed. On the contrary, for the two configurations of FePc on Ag(111) the charge redistribution is spread out across the whole molecule showing both, the metal center and the aromatic macrocycle (including the outermost carbon rings), regions of increased/decreased electron density. This relatively strong charge transfer amplifies the symmetry-breaking through charge redistribution effects already present due to adsorption, leading to additional splitting of $E_g$ modes and mode-specific frequency shifts which may further explain the observed discrepancies between our submolecularly-resolved TERS spectra and the Raman spectra obtained previously on FePc thin films.[33,34]

Our results have implications on the nascent field of chemical recognition of molecules with submolecularly resolved hyperspectral TERS maps. We demonstrate that it is possible to detect the adsorption-induced symmetry breaking and detect sub-ångstrom structural displacements by experimental optical means.[35] We anticipate that this phenomenon will be also observable for molecules undergoing Jahn-Teller distortions.[36] We also show that the effect of the substrate and the resulting overall chirality can severely alter the expected optical properties at the picoscale,[37-39] and that the symmetry of the substrate must be considered when reconstructing the chemical structure of an unknown adsorbate from TERS measurements. We anticipate future theoretical work on the interpretation and simulation of the substrate-induced distortions of Ångstrom-resolved Raman maps.[40]

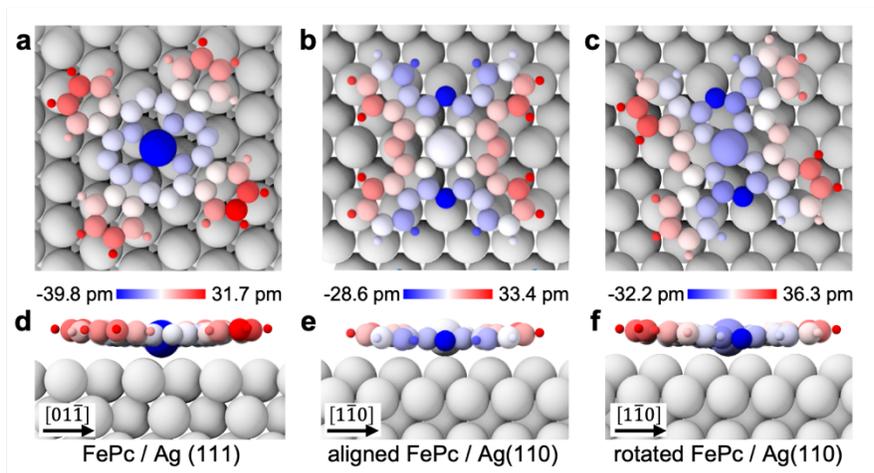

*Figure 4.* Adsorption-induced out-of-plane relaxations and symmetry breaking calculated by density functional theory calculations. **a-c.** Front view of the optimized geometries for the FePc/Ag(111), aligned FePc/Ag(110) and rotated FePc/Ag(110) configuration respectively. **d-f.** Side views of the optimized FePc/Ag(111), aligned FePc/Ag(110) and rotated FePc/Ag(110) geometries respectively. The height (z) atomic displacements are color-coded with respect to the average z-coordinate of the molecule: atoms located below the mean height are shown in blue, while those above it are shown in red (closer to or further from the Ag surface, respectively).

In conclusion, we have characterized vibrational modes of individual FePc molecules adsorbed on two low-index Ag crystal surface terminations by means of TERS with submolecular resolution. The vibrational mode patterns were successfully analyzed for three adsorption configurations that represent a gradual reduction of the original $D_{4h}$ symmetry of the molecule to $C_{2d}$, $C_{2v}$ and $C_2$ symmetries, depending on surface termination and the molecular in-plane rotational angle on it. We have demonstrated that the symmetry breaking has a strong impact on the observed hyperspectral patterns, proving that TERS is sensitive to picometer-scale distortions of the position of the atoms of the adsorbates. We have identified the vibrational modes for which the degeneracy has been lifted, revealing how subtle variations in the adsorption geometry and registry with the underlying substrate profoundly affect the activity of the molecular vibrations, of relevance to control on-surface reactions with enhanced chemoselectivity. Our work creates a basis for precisely studying point-group symmetries of single molecules by using purely optical methodologies.

**Author contribution.**

R.C.C.F and B.C. contributed equally.

**Supplementary Information.**

The Supporting Information is available free of charge at XXX

Methods, Experimental details, extended microspectroscopic data and extended calculations.

**Acknowledgements**.

We acknowledge fruitful discussions with Dr. Tomáš Neuman from the Institute of Physics. The authors acknowledge financial support from the Spanish MICIN (grants PID2021-12509OA-I00, PID2022-138470NB-I00, TED2021-129416A-I00, PID2024-155345NA-I00, RYC2024-049194-I, and CNS2022-135658 funded by MCIN/AEI/10.13039/501100011033 and by "ERDF A way of making Europe", by "ERDF/EU" and the "European Union NextGenerationEU/ PRTR") and TEC-2024/TEC-459 project funded by the "Comunidad de Madrid" and co-financed by European Structural Funds. The authors acknowledge the Severo Ochoa Centres of Excellence program through Grant CEX2024-001445-S. BC acknowledges support from the Spanish Comunidad de Madrid "Talento Program César Nombela" (project No. 2023-T1/TEC-28968). R.C.C.F, A.S., P.K. and M.Š. acknowledge the Czech Science Foundation grant no. 22–18718S and the support from the CzechNanoLab Research Infrastructure supported by MEYS CR (LM2023051). All authors are grateful to Mobility Plus bilateral grant no. BILAT23033/CSIC-24–10 from the Spanish CSIC and the Czech Academy of Sciences. We acknowledge the allocation of computing time at the Centro de Computación Científica at the Universidad Autónoma de Madrid (CCC-UAM)